\documentclass[ floatfix,twocolumn]{revtex4}
\usepackage{amsmath}
\usepackage{color}
\usepackage{amssymb}

\usepackage{graphicx}

\begin{document}

{\bf ~~\\This article is devoted to the 100th birthday \\
of my teacher Isaak Markovich Khalatnikov\\~~\\~~ }
    
\title{ Low-Temperature Transport  in Metals without Inversion Centre }

\author{V.P.Mineev$^{1,2\footnote{E-mail: vladimir.mineev@cea.fr}}$}
\affiliation{$^1$Univ. Grenoble Alpes, CEA, IRIG, PHELIQS, F-38000 Grenoble, France\\
$^2$Landau Institute for Theoretical Physics, 142432 Chernogolovka, Russia}

\begin{abstract}
Theory of low temperature kinetic phenomena in metals  without inversion center is developed.
Kinetic properties of a metal without inversion center
are described by   four kinetic equations
for the  diagonal (intra-band)  and the off-diagonal (inter-band)  elements of matrix distribution function of electrons occupying the states in two bands split by the spin-orbit interaction. 
The derivation of collision integrals for electron-impurity scatterings and for electron-electron scatterings in a non-centrosymmetric medium is given. 
Charge, spin and heat transport in 
the ballistic and the weak impurity scattering regimes is discussed.
It is shown that the off-diagonal terms give rise the contribution in  charge, spin  and heat flows   not only due to the interband scattering 
but also  
 in the  collisionless case.
The zero-temperature residual resistivity and the residual thermal resistivity are determined 
by scattering on impurities as well as by the electron-electron scattering. 
\end{abstract}

\date{\today}
\maketitle

\section{Introduction}

During the last decades or so, there has been great interest to the spin-based electronics in application to the  systems where spin-orbit coupling plays an important role. In particular there were studied the kinetic properties of two-dimensional semiconductors with broken space parity characterized by  both Rashba and Dresselhaus interaction\cite{Mishchenko2003,Loss2003,Huang2006,Grimaldi2016}. Spin-orbit interaction of electrons with a non-centrosymmetric crystal lattice lifts spin degeneracy of electron states.
Each band filled by twice degenerate electron states splits on two bands filled by the electron states with different momenta at the same energy. Usually scalar electron energy and the Fermi distribution function are given  now   by the matrices $\hat\varepsilon=\varepsilon_{\sigma\sigma^\prime}$,  $\hat n=n_{\sigma\sigma^\prime}$ in respect of  spin indices.
The distribution matrix time variation in the $({\bf k},{\bf r})$ space is determined by the quasi-classic kinetic equation that was derived by V.P.Silin
\cite{Silin1957} and has the following form
\begin{eqnarray}
\frac{\partial \hat n}{\partial t}+\frac{1}{2}\left ( \frac{\partial\hat \varepsilon}{\partial {\bf k}}
\frac{\partial\hat n}{\partial {\bf r}} +\frac{\partial\hat n}{\partial {\bf r}} \frac{\partial\hat \varepsilon}{\partial {\bf k}}  \right )-
\frac{1}{2}\left ( \frac{\partial\hat \varepsilon}{\partial {\bf r}}
\frac{\partial\hat n}{\partial {\bf k}} +\frac{\partial\hat n}{\partial {\bf k}} \frac{\partial\hat \varepsilon}{\partial {\bf r}}  \right )\nonumber\\-i[\hat\varepsilon,\hat n]=\hat I_{st},~~~~~~~~~~~~~~~~~~~~~~~~
\label{Silin}
\end{eqnarray}
where $[\hat\varepsilon,\hat n]$ is the commutator of $\hat\varepsilon=\hat\varepsilon({\bf k}, {\bf r})$ and $\hat n=\hat n({\bf k},{\bf r})$. We put $\hbar=1$. The  collision integral in the rhs determines the relaxation processes. 

It is quite natural to rewrite kinetic equation in the band representation where the Hamiltonian is diagonal. It seems that after this transformation we  come to   kinetic equations for distribution functions in each  band
which interact each other  due to collision integrals including inter-band scattering. So, the theory seems to be similar to the kinetic theory of a two band metal with center of inversion. 
However,  this is not the case.  Kinetic processes in a  non-centrosymmetric medium  are described  by   four kinetic equations
for the  diagonal (intra-band)  and the off-diagonal (inter-band)  elements of matrix distribution function of electrons occupying the states in two bands split by the spin-orbit interaction.
The off-diagonal terms give rise the contribution in the transport properties even in the collisionless regime.

The expression for the electron-impurity collision integral in non-centrosymmetric semiconductors or metals one can find in the papers \cite{Koshelev1988,Khaetskii2006}. The authors do not derive the collision integral but write: "the collision term was derived in many papers" and give the corresponding references.   These references, however, do not contain a  derivation of collision integral. 
The derivation of the collision integrals for the electron-impurity collisions as well as for electron-electron collisions  in a non-centrosymmetric medium is given in the present article.
Along with scattering on impurities the electron-electron collisions in non-centrosymmetric medium
leads to the  zero-temperature residual resistivity  and residual thermal conductivity.

The paper is organized as follows. Section II contains the basic notions of the electron energy spectrum and the equilibrium distribution in metals without inversion. In the Section III  there are presented the system of kinetic equations 
and derived 
the  expressions for  electric current, spin current and heat current.  For each type of current the collisionless regime and the 
 weak impurity scattering case are examined. There is shown that the off-diagonal terms give rise the contribution in  charge, spin  and heat transport   not only due to interband scattering 
but also  
 in the  collisionless case.
The role of electron-electron scattering in formation of zero-temperature residual resistivity and residual thermal conductivity is discussed in the Section IV. In the Conclusion there are enumerated the principal results of the paper.
The derivations
of collision integral for the electron scattering on scalar impurities as well as  for electron-electron scattering are given  in the Appendices A and B. The results are derived in application to a medium without inversion center both in two and three dimensional case.

\section{Electronic states in non-centrosymmetric metals}

 The spectrum of noninteracting
electrons in a metal without inversion center is:
\begin{equation}
\label{H_0}
 \hat \varepsilon({\bf k})
 = \varepsilon({\bf k})\hat\delta+\mbox{\boldmath$\gamma$}({\bf k}) 
   \cdot \mbox{\boldmath$\sigma$},
\end{equation}
where 
$\varepsilon({\bf k})$ denotes the spin-independent part of the spectrum ,
$\hat\delta$ is the unit $2\times 2$ matrix in the spin space, $\mbox{\boldmath$\sigma$}=(\sigma_x,\sigma_y,\sigma_z)$ are the Pauli matrices. 
The second term in Eq.
(\ref{H_0}) describes the  spin-orbit  coupling whose form depends on the specific noncentrosymmetric crystal structure.
The pseudovector $\mbox{\boldmath$\gamma$}({\bf k})$  satisfies
$\mbox{\boldmath$\gamma$}(-{\bf k})=-\mbox{\boldmath$\gamma$}({\bf k})$ and 
$g\mbox{\boldmath$\gamma$}(g^{-1} {\bf k})=\mbox{\boldmath$\gamma$}({\bf k})$,
where $g$ is any symmetry operation in the  point group ${\cal G}$ of
the crystal. A more detailed  theoretical description of noncentrosymmetric metals in normal and in superconducting state is presented in the paper \cite{Mineev2012}. 
The tetragonal point group
${\cal G}=\mathbf{C}_{4v}$, relevant for CePt$_3$Si,
CeRhSi$_3$ and CeIrSi$_3$, yields the antisymmetric spin-orbit coupling 
\begin{equation}
\label{gammaC4v}
\mbox{\boldmath$\gamma$}({\bf k})=\gamma(k_y\hat x-k_x\hat y)
    +\gamma_\parallel k_xk_yk_z(k_x^2-k_y^2)\hat z.
\end{equation}
In the purely two-dimensional case, setting $\gamma_\parallel=0$
one recovers the Rashba interaction \cite{Rashba1960} which is often
used to describe the effects of the absence of mirror symmetry in
semiconductor quantum wells.
The case of isotropic spectrum when $\varepsilon({\bf k})=\frac{k^2}{2m}$ and 
\begin{equation}
\label{gammaO}
    \mbox{\boldmath$\gamma$}({\bf k})=\gamma{\bf k}
\end{equation}
is compatible with the 3D cubic crystal symmetry. Here $\gamma$ is a constant. 

The eigenvalues  and eigenfunctions of the matrix (\ref{H_0}) are
\begin{equation}
    \varepsilon_{\pm}({\bf k})=\varepsilon({\bf k})\pm
    |\mbox{\boldmath$\gamma$}({\bf k})|,
\label{e3}
\end{equation}
\begin{eqnarray}
\Psi^+_\sigma({\bf k})=C_{\bf k}\left (\begin{array} {c}
\hat\gamma_{{\bf k}z}+1\\
\hat\gamma_{{\bf k}x}+i\hat\gamma_{{\bf k}y}
\end{array}\right),\nonumber\\
~~~~~~~~~~~~\Psi^-_\sigma({\bf k})=C_{\bf k}\left(\begin{array} {c}
-\hat\gamma_{{\bf k}x}+i\hat\gamma_{{\bf k}y}\nonumber\\
\hat\gamma_{{\bf k}z}+1
\end{array}\right),\\
~~~~~~~~~~C_{\bf k}=(2(\hat\gamma_{{\bf k}z}+1))^{-1/2}.~
\end{eqnarray}
Here, $\hat\gamma_{{\bf k}x},\hat\gamma_{{\bf k}y},\hat\gamma_{{\bf k}z}$ are the components of the unit vector
$\mbox{\boldmath$\gamma$}({\bf k})/|\mbox{\boldmath$\gamma$}({\bf k})|$. The eigen functions obey the orthogonality conditions
\begin{equation}
\Psi^{\alpha\star}_\sigma({\bf k})\Psi^\beta_\sigma({\bf k})=\delta_{\alpha\beta},~~~~~~~
\Psi^\alpha_{\sigma_1}({\bf k})\Psi^{\alpha\star}_{\sigma_2}({\bf k})=\delta_{\sigma_1\sigma_2}.
\label{ort}
\end{equation}
Here, and in all the subsequent formulas there is implied the summation over the repeating  spin $\sigma=\uparrow,\downarrow$
or band $\alpha=+,-$ indices.

There are two Fermi surfaces determined by the equations
\begin{equation}
\label{e4}
    \varepsilon_{\pm}({\bf k})=\mu
\end{equation}
with different Fermi momenta ${\bf k}_{F\pm}$. In the Rashba 2D model and in the 3D isotropic case they are
\begin{equation}
k_{F\pm}=\mp m \gamma+\sqrt{2m\mu+(m\gamma)^2}
\end{equation}
and the Fermi velocity  has the  common value 
\begin{equation}
{\bf v}_{F\pm}=\frac{\partial(\varepsilon \pm\gamma k)}{\partial {\bf k}}|_{k=k_{F\pm}}=\hat{\bf k}\sqrt{\frac{2\mu}{m}+\gamma^2},
\end{equation}
here $\hat{\bf k}$ is the unit vector along momentum ${\bf k}$. The equivalence of the Fermi velocities at different Fermi momenta is the particular property of the models with isotropic spin-orbital coupling (\ref{gammaO}) in 3D case and the Rashba interaction in 2D case.

The matrix of equilibrium electron distribution function is
\begin{equation}
\hat n^0
=\frac{n_++n_-}{2}\hat\delta+\frac{n_+-n_-}{2|\mbox{\boldmath$\gamma$}|} \mbox{\boldmath$\gamma$} \cdot \mbox{\boldmath$\sigma$},
\label{eqv}
\end{equation}
where 
\begin{equation}
n_\pm=\frac{1}{\exp\left(\frac{\varepsilon_{\pm}-\mu}{T}\right)+1}
\end{equation}
are the Fermi functions.
In the isotropic case near the corresponding Fermi surfaces the dispersion laws  have the particular simple form
\begin{equation}
\xi_\pm=\varepsilon_\pm-\mu\approx v_F(k-k_{F\pm})=\epsilon-\mu_\pm,
\end{equation}
with 
\begin{equation}
\epsilon=v_Fk,~~~~~\mu_\pm=v_Fk_{F\pm},~~~~~~\mu_+-\mu_-=-2mv_F\gamma.
\label{ene}
\end{equation}

\section{Trasport properties. Impurity scattering.}
\subsection{Kinetic equation}

In presence of time dependent  electric field ${\bf E}(t)={\bf E}_{\omega}e^{-i\omega t}$ the linearized kinetic equation (\ref{Silin}) is
\begin{eqnarray}
\frac{\partial \hat g}{\partial t}+e{\bf E}
\frac{\partial\hat n^0}{\partial {\bf k}}-i[\hat\varepsilon,\hat g]
=\hat I_{st},
\label{Silin1}
\end{eqnarray}
where $\hat g=\hat n-\hat n^0$ is the deviation of distribution function from equilibrium distribution $\hat n^0$.

The hermitian matrices of the nonequilibrium distribution functions in band and spin representations are related as 
\begin{equation}
f_{\alpha\beta}({\bf k})=\Psi^{\alpha\star}_{\sigma_1}({\bf k})n_{\sigma_1\sigma_2}\Psi^{\beta}_{\sigma_2}({\bf k}).
\end{equation}
In the band representation the equilibrium distribution function (\ref{eqv}) is the diagonal matrix
\begin{equation}
f_{\alpha\beta}^0({\bf k})=\Psi^{\alpha\star}_{\sigma_1}({\bf k})n^0_{\sigma_1\sigma_2}\Psi^{\beta}_{\sigma_2}({\bf k})=\left (\begin{array} {cc} n_+&0\\0&n_-  \end{array}\right)_{\alpha\beta}.
\end{equation}
However, the matrix of derivative of the equilibrium distribution in the band representation is not diagonal and given by the following equation
\begin{eqnarray}
\Psi^{\alpha\star}_{\sigma_1}({\bf k})\frac{\partial n^0_{\sigma_1\sigma_2}}{\partial {\bf k}}\Psi^{\beta}_{\sigma_2}({\bf k})=
\frac{\partial f^0_{\alpha\beta}}{\partial {\bf k}}+\left [ \Psi^{\alpha\star}_{\sigma}({\bf k}) \frac{\partial\Psi^{\gamma}_{\sigma}({\bf k})}{\partial {\bf k}}, f^0_{\gamma\beta} \right],~~
\end{eqnarray}
where $\left [\dots,\dots\right ]$ is the commutator.
Hence, the matrix  kinetic equation  for the  frequency dependent Fourier amplitudes of non-equilibrium part of distribution function $g_{\alpha\beta}({\bf k},t)=g_{\alpha\beta}({\bf k},\omega)e^{-i\omega t}$ acquires the form
\begin{widetext}
\begin{eqnarray}
-i\omega
 \left (\begin{array} {cc}g_+&g_{\pm}\\g_{\mp}&g_-
 \end{array}\right)
 +e\left (\begin{array} {cc}({\bf v}_{+}{\bf E}) \frac{\partial n_+}{\partial \xi_+}&({\bf v}_{\pm}{\bf E})(n_--n_+)\\
({\bf v}_{\mp}{\bf E})(n_+-n_-)&({\bf v}_{-}{\bf E})\frac{\partial n_-}{\partial \xi_-}
 \end{array}\right)+
 \left(
\begin{array} {cc}0&ig_{\pm}(\varepsilon_--\varepsilon_+)\\
ig_{\mp}(\varepsilon_+-\varepsilon_-)&0
 \end{array}\right)=I_{\alpha\beta}
 \label{eqv1}
\end{eqnarray}
Here 
\begin{equation}
{\bf v}_\alpha=\frac{\partial\varepsilon_\alpha}{\partial{\bf k}},~~~~~{\bf v}_{\pm}=
\Psi^{+\star}_{\sigma}({\bf k})\frac{\partial \Psi^{-}_{\sigma}({\bf k})}{\partial{\bf k}}=\frac{(\hat\gamma_{{\bf k}x}-i\hat\gamma_{{\bf k}y})}{2(\hat\gamma_{{\bf k}z}+1)}
\frac {\partial \hat\gamma_{{\bf k}z}}{\partial{\bf k}}
-\frac{1}{2}
\frac {\partial(\hat\gamma_{{\bf k}x}-i\hat\gamma_{{\bf k}y})
}{\partial{\bf k}},~~~~{\bf v}_{\mp}=-{\bf v}_{\pm}^\star
\label{vel}
\end{equation}
The collision integral $I_{\alpha\beta}$ for electron scattering on impurities is derived in Appendix A. One can check that it is equal to zero in  
equilibrium. Hence, it is
\begin{eqnarray}
I_{\alpha\beta}({\bf k})=2\pi n_{imp}\int\frac{d^3k^\prime}{(2\pi)^3}|V({\bf k}-{\bf k}^\prime)|^2\left \{O_{\alpha\nu}({\bf k},{\bf k}^\prime)\left [ g_{ \nu\mu}({\bf k}^\prime)O_{\mu\beta}({\bf k}^\prime,{\bf k})-O_{\nu\mu}({\bf k}^\prime,{\bf k})
  g_{ \mu\beta}({\bf k}) \right ]\delta(\varepsilon^\prime_\nu-\varepsilon_\beta)\right.\nonumber\\ 
 \left .+
  \left[O_{\alpha\nu}({\bf k},{\bf k}^\prime)g_{ \nu\mu}({\bf k}^\prime)-g_{ \alpha\nu}({\bf k})O_{\nu\mu}({\bf k},{\bf k}^\prime)
  \right ]O_{\mu\beta}({\bf k}^\prime,{\bf k})\delta(\varepsilon^\prime_\mu-\varepsilon_\alpha)\right \},
  \label{matrix1}
\end{eqnarray}
\end{widetext}
\begin{equation}
O_{\alpha\beta}({\bf k},{\bf k}^\prime)=\Psi^{\alpha\star}_\sigma({\bf k})\Psi^\beta_\sigma({\bf k}^\prime).
\end{equation}
Here, and in all the subsequent equations when we will discuss 2D case one must substitute the 3D integration over  reciprocal space 
$\int\frac{d^3k^\prime}{(2\pi)^3}$
by the corresponding 2D expression $\int\frac{d^2k^\prime}{(2\pi)^2}$.

The solution of Eq.(\ref{eqv1}) 
 has the following form
\begin{equation}
g_{\alpha\beta}= \left (\begin{array} {cc}g_+&g_{\pm}\\g_{\mp}&g_-
 \end{array}\right)=
 e\left (\begin{array} {cc}({\bf w}_+{\bf E})
 &({\bf w}_\pm{\bf E})
\\
({\bf w}_\mp{\bf E})
&({\bf w}_-{\bf E})
 \end{array}\right).
 \label{sol}
 \end{equation}
After substitution this matrix 
in the Eq.(\ref{eqv1}) and in the collision integral Eq.(\ref{matrix1}) we obtain  
 four scalar equations corresponding to each matrix element
of  the matrix Eq.(\ref{eqv1}) for ${\bf k} $ dependent four scalar functions   $ ({\bf w}_+{\bf E}),  
({\bf w}_\pm {\bf E}),
 ({\bf w}_\mp{\bf E}), 
 ({\bf w}_- {\bf E})$. These functions,  in general,  can be determined   by solving the equations numerically. The particular solutions for collisionless regime and the weak impurity scattering case are considered in the next sections.
 
\subsection{Electric current}
 
  The electric  current density is
 \begin{equation}
 {\bf j}=e\int \frac{d^3k}{(2\pi)^3}\frac{\partial \varepsilon_{\sigma\sigma_1}({\bf k})}{\partial {\bf k}}g_{\sigma_1\sigma}({\bf k},\omega)
 \end{equation}
Transforming it to the band representation we obtain
\begin{widetext}
\begin{eqnarray}
 {\bf j}=e\int \frac{d^3k}{(2\pi)^3} \Psi_\sigma^{\alpha\star}({\bf k})\frac{\partial \varepsilon_{\sigma\sigma_1}({\bf k})}{\partial {\bf k}}
 \Psi_{\sigma_1}^{\gamma}({\bf k})\Psi_{\sigma_2}^{\gamma\star}({\bf k})
 g_{\sigma_2\sigma_3}({\bf k},\omega)\Psi_{\sigma_3}^\alpha({\bf k})\nonumber\\
 =e\int \frac{d^3k}{(2\pi)^3} \left \{
 \frac{\partial \varepsilon_{\alpha\gamma}({\bf k})}{\partial {\bf k}}
 +\left [ \Psi_{\sigma}^{\alpha\star}({\bf k})
 \frac{\partial\Psi^{\beta}_{\sigma}}{\partial {\bf k}},  \varepsilon_{\beta\gamma} \right]\right\}
 g_{\gamma\alpha}({\bf k},\omega),
\end{eqnarray}
\end{widetext}
where $\left [\dots,\dots\right ]$ is the commutator.
Finally we come to
\begin{eqnarray}
 {\bf j}=e^2\int \frac{d^3k}{(2\pi)^3} \left\{
{\bf v}_+({\bf w}_+{\bf E})
+
{\bf v}_-({\bf w}_-{\bf E})\right.\nonumber\\
\left.+
\left[{\bf v}_\pm({\bf w}_\mp{\bf E})-
 {\bf v}_\mp({\bf w}_\pm{\bf E})
\right ](\varepsilon_--\varepsilon_+)  \right \}.
\label{current}
\end{eqnarray}
 The functions $ {\bf w}_+,{\bf w}_- ,  
{\bf w}_\pm,
 {\bf w}_\mp
 $ depend from the modulus and the direction of momentum ${\bf k}$. Because of this the direction of electric current does coincide 
 in general with the direction of electric field.
 
 \subsubsection{Ballistic regime} 
 
 In neglect the scattering terms, that is at $\omega\tau>1$ where $\tau$ is the symbol for the typical times of scattering determined by the different terms in the scattering integral,
  the Eq.(\ref{eqv1}) has the following solution
\begin{eqnarray}
g_+=e({\bf w}_+{\bf E})=\frac{e}{i\omega}({\bf v}_{+}{\bf E}) \frac{\partial n_+}{\partial \xi_+},
\label{A}\\
g_-=e({\bf w}_-{\bf E})=\frac{e}{i\omega}({\bf v}_{-}{\bf E}) \frac{\partial n_-}{\partial \xi_-},
\label{B}\\
g_{\pm}=e({\bf w}_{\pm}{\bf E})=\frac{e({\bf v}_{\pm}{\bf E})(n_--n_+)}{i\omega-i(\varepsilon_--\varepsilon_+)},
\label{C}\\
g_{\mp}=e({\bf w}_{\mp}{\bf E})=\frac{e({\bf v}_{\mp}{\bf E})(n_+-n_-)}{i\omega-i(\varepsilon_+-\varepsilon_-)}.
\label{D}
\end{eqnarray}
Substitution these expressions to the Eq.(\ref{current}) gives
\begin{eqnarray}
 {\bf j}=e^2\int \frac{d^3k}{(2\pi)^3} \left\{
\frac{{\bf v}_+({\bf v}_{+}{\bf E})}{i\omega} \frac{\partial n_+}{\partial \xi_+}
+
\frac{{\bf v}_-({\bf v}_{-}{\bf E})}{i\omega} \frac{\partial n_-}{\partial \xi_-}
\right.\nonumber\\
\left.+2\frac{(n_+-n_-)(\varepsilon_--\varepsilon_+)}{\omega^2-(\varepsilon_+-\varepsilon_-)^2}
\left [i\omega\mbox{Re(}{\bf v}_\pm({\bf v}_\pm^\star{\bf E}))\right.\right.\nonumber\\
+\left.\left.(\varepsilon_--\varepsilon_+)]\mbox{Im}({\bf v}_\pm({\bf v}_\pm^\star{\bf E}))\right ]
 \right \}.
\label{cur1}
\end{eqnarray}
The last term in this formula is in fact equal to zero because
the combination
\begin{eqnarray}
\mbox{Im}(v_{\pm i}v_{\pm j}^\star) E_j=\frac{1}{2}\mbox{\boldmath$\hat\gamma$}\left (\frac{\partial\mbox{\boldmath$\hat\gamma$}}{\partial k_i}\times
 \frac{\partial\mbox{\boldmath$\hat\gamma$}}{\partial k_j}\right )E_j,
\end{eqnarray}
in 2D case  is equal to zero and in 3D case it is odd function of ${\bf k}$. 
 
The integral in the last two terms of Eq.(\ref{cur1}) is taken over the region of reciprocal space in between $k_{F+}$ and $k_{F-}$, where $\varepsilon_+-\varepsilon_-\approx 2\gamma k_F$. At relatively small frequencies $\omega\ll 2\gamma k_F$, but still in collisionless regime $\omega>
\tau^{-1}$ one can rewrite the Eq.(\ref{cur1}) as
\begin{eqnarray}
 {\bf j}=e^2\int \frac{d^3k}{(2\pi)^3} \left\{
\frac{{\bf v}_+({\bf v}_{+}{\bf E})}{i\omega} \frac{\partial n_+}{\partial \xi_+}
+
\frac{{\bf v}_-({\bf v}_{-}{\bf E})}{i\omega} \frac{\partial n_-}{\partial \xi_-}
\right\}.
\label{cur2}
\end{eqnarray}

\subsubsection{Weak impurity scattering}

The weak impurity scattering regime is limited by inequality $\tau^{-1}\ll 2\gamma k_F$. This case one can neglect by the scattering terms in the kinetic equations for the off-diagonal elements of distribution function.
Thus, the solutions for off-diagonal matrix elements of distribution function still is given by Eqs.(\ref{C}) and (\ref{D}). After substitution of these solutions in the collision integral in the equations for the diagonal elements of distribution function we come to the equations
\begin{widetext}
\begin{eqnarray}
-i\omega g_++e({\bf v}_+{\bf E}))\frac{\partial n_+}{\partial \xi_+}=
4\pi n_i\int\frac{d^3k}{2\pi^3}|V({\bf k}-{\bf k}^\prime)|^2\times\nonumber~~~~~~~~~~~~~~~~~~~~~~~~~~~~~~~~~~~~~~~~~~~~~~
\\ \times\left\{
O_{++}({\bf k}{\bf k}^\prime)O_{++}({\bf k}^\prime{\bf k})[g_+({\bf k}^\prime)-g_+({\bf k})]\delta(\varepsilon_+^\prime-\varepsilon_+)+
O_{+-}({\bf k}{\bf k}^\prime)O_{-+}({\bf k}^\prime{\bf k})[g_-({\bf k}^\prime)-g_+({\bf k}))]delta(\varepsilon_-^\prime-\varepsilon_+)
\right \},~~\label{34}\\
-i\omega g_-+e({\bf v}_-{\bf E}))\frac{\partial n_-}{\partial \xi_-}=4\pi n_i\int\frac{d^3k}{2\pi^3}|V({\bf k}-{\bf k}^\prime)|^2\times\nonumber~~~~~~~~~~~~~~~~~~~~~~~~~~~~~~~~~~~~~~~~~~~~~~
\\
\times\left\{
O_{-+}({\bf k}{\bf k}^\prime)O_{+-}({\bf k}^\prime{\bf k})[g_+({\bf k}^\prime)-g_-({\bf k})]\delta(\varepsilon_+^\prime-\varepsilon_-)+
O_{--}({\bf k}{\bf k}^\prime)O_{--}({\bf k}^\prime{\bf k})[g_-({\bf k}^\prime)-g_-({\bf k})]\delta(\varepsilon_-^\prime-\varepsilon_-)
\right \},~~\label{35}
\end{eqnarray}
\end{widetext}
obtained  neglecting  in the scattering integral by the  terms  with off-diagonal components of the distribution function. These terms  are $\gamma k_F\tau \gg1$ times smaller than the terms with  diagonal elements. We see that even in the limit of weak impurity scattering
the relaxation of diagonal elements of distribution function to equilibrium is determined  in general by the four different collision terms. 

There were undertaken several attempts \cite{Loss2003,Huang2006, Grimaldi2016} to solve these equations for the 2D Rashba model in the Born approximation.
This case the products
$$
O_{++}({\bf k}{\bf k}^\prime)O_{++}({\bf k}^\prime{\bf k})
=O_{--}({\bf k}{\bf k}^\prime)O_{--}({\bf k}^\prime{\bf k})=\cos^2\frac{\varphi-\varphi^\prime}{2},$$
$$O_{+-}({\bf k}{\bf k}^\prime)O_{-+}({\bf k}^\prime{\bf k})
=O_{-+}({\bf k}{\bf k}^\prime)O_{+-}({\bf k}^\prime{\bf k})=-\sin^2\frac{\varphi-\varphi^\prime}{2}~$$ depend from the difference of the azimuthal angles of 
initial and final vector of momentum.
In the Born approximation the scattering integral is expressed through the  potential of scattering depending from transferred momentum
$V({\bf k}-{\bf k}^\prime)$ that  means it also depends from $\varphi-\varphi^\prime$. This creates possibility to search 
  the solution of Eqs. (\ref{34}), (\ref{35})
  in the following form 
 \begin{eqnarray}
g_+=-ea_+\frac{\partial n_+}{\partial \xi_+}({\bf v}_+{\bf E}),~~~
g_-=-ea_-\frac{\partial n_-}{\partial \xi_-}({\bf v}_-{\bf E})
,
\label{sol}
\end{eqnarray}
as it was done in Ref.4 where the coefficients $a_+,a_-$ were found at $\omega=0$.
Then the  current is 
\begin{eqnarray}
 {\bf j}=e^2\int \frac{d^2k}{(2\pi)^2} \left\{
a_+{\bf v}_+({\bf v}_{+}{\bf E}) \frac{\partial n_+}{\partial \xi_+}
+
a_-{\bf v}_-({\bf v}_{-}{\bf E}) \frac{\partial n_-}{\partial \xi_-}\right\}.~~
\label{cur3}
\end{eqnarray}
 This approach easily generalized to the finite frequency case.

 Similar treatment is possible for the Dresselhaus model \cite{footnote} where $\mbox{\boldmath$\gamma$}({\bf k})=\gamma_D(k_y\hat y-k_x\hat x)$, but not for the model where vector $\mbox{\boldmath$\gamma$}({\bf k})$
is given by the sum of vectors in the Rashba and the Dresselhaus models.

 For  other 2D or 3D models one can search the solution of Eqs.(\ref{34}) and (\ref{35}) by  numerical methods. 
 In weak impurity scattering regime the current is determined as
 \begin{eqnarray}
  {\bf j}=e^2\int \frac{d^3k}{(2\pi)^3} \left\{
{\bf v}_+({\bf w}_+{\bf E})
+
{\bf v}_-({\bf w}_-{\bf E})
 \right \}.~~~~~~~~~~~~~
 \end{eqnarray}

 \subsubsection{Regime of strong scattering}
 
 The strong scattering occurs when the typical inverse scattering time is of the order of spin-orbit band splitting $\tau^{-1}\approx \gamma k_F$. This case at $\gamma k_F\ll\varepsilon_F$ the quasi-classical kinetic theory is still applicable to description of the kinetic phenomena but one must solve the whole system kinetic equations for the diagonal and the off-diagonal matrix elements of the distribution function.

\subsection{Spin current}

An electric field in a crystal without inversion center generates a spin current.
The  density of spin current arising in an electric field is
 \begin{equation}
 {\bf j}_{i}=e\int \frac{d^3k}{(2\pi)^3}\mbox{\boldmath$\sigma$}_{\sigma\sigma_1}\frac{\partial \varepsilon_{\sigma_1\sigma_2}({\bf k})}{\partial k_i}g_{\sigma_2\sigma}({\bf k},\omega).
\end{equation}
Transforming it to the band representation in the same manner as it was done for electric current we come to the following expressions for the spin current components
\begin{widetext}
\begin{eqnarray}
 j_{xi}=e\int \frac{d^3k}{(2\pi)^3} \left\{
\left [{v}_{+i}({\bf w}_\pm{\bf E})
+{v}_{-i}({\bf w}_\mp{\bf E})\right ]
+
\left[ v_{\pm i}({\bf w}_-{\bf E})
-
  v_{\mp i}({\bf w}_+{\bf E})
 \right ]
(\varepsilon_--\varepsilon_+)  \right \}
,\label{sc1}\\
 j_{yi}=
 ie\int \frac{d^3k}{(2\pi)^3} \left\{
\left [{v}_{+i}({\bf w}_\pm{\bf E})
-{v}_{-i}({\bf w}_\mp{\bf E})\right ]
+
\left[v_{\pm i}({\bf w}_-{\bf E})
+
  v_{\mp i}({\bf w}_+{\bf E})
 \right ]
(\varepsilon_--\varepsilon_+)  \right \}
,\label{sc2}\\
  j_{zi}=
 e\int \frac{d^3k}{(2\pi)^3} \left\{
{v}_{+i}({\bf w}_+{\bf E})
-{v}_{-i}({\bf w}_-{\bf E})
+
\left[v_{\pm i}({\bf w}_\mp{\bf E})+
 v_{\mp i}({\bf w}_\pm{\bf E})\right ]
(\varepsilon_--\varepsilon_+)  \right \}
 .
 \label{sc3}
\end{eqnarray}
\end{widetext}
In collisionless regime the solutions of kinetic equation are given by Eqs.(\ref{A})-(\ref{D}).
In two-dimensional case the velocities Eq.(\ref{vel})  are
\begin{equation}
{\bf v}_\alpha=\frac{\partial\varepsilon_\alpha}{\partial{\bf k}},~~~{\bf v}_{\pm}=
-\frac{1}{2}\frac {\partial(\hat\gamma_{{\bf k}x}-i\hat\gamma_{{\bf k}y})}{\partial{\bf k}}.
\label{vel2}
\end{equation}
Obviously, the "diagonal" velocities are odd functions of the wave vector ${\bf v}_\alpha(-{\bf k})=-{\bf v}_\alpha({\bf k})$,
and the "off-diagonal" velocities are even functions of the wave vector ${\bf v}_{\pm}(-{\bf k})={\bf v}_\pm({\bf k})$.
Hence,  in a 2D non-centrosymmetric media 
 the ballistic spin currents Eqs.(\ref{sc1}), (\ref{sc2})
are identically equal to zero  
 \begin{equation}
  j_{xi}= j_{yi}=0. 
  \label{sc}
 \end{equation}
 
 To calculate the spin current
 in the case of weak impurity scattering, one can use  the off-diagonal matrix elements of distribution function given by Eqs.(\ref{C})-(\ref{D}) but for  the diagonal elements one should solve the equations (\ref{34})-(\ref{35}).  For the 2D Rashba model in the Born approximation the solution is given by the Eq.(\ref{sol}). Thus, in this case the spin current is also equal to zero due to the parity properties of "diagonal" and "off-diagonal" velocities.

In three dimensions the "off-diagonal" velocities Eq.(\ref{vel})
are not even functions any more. The spin current  under electric field acquire finite value.

\subsection{Heat current}

In  presence of temperature gradient  the matrix  kinetic equation  for 
non-equilibrium distribution function $g_{\alpha\beta}$ is
\begin{widetext}
\begin{eqnarray}
 -\frac{1}{T}\left(
\begin{array} {cc} ({\bf v}_{+}\nabla T)\xi_+ \frac{\partial n_+}{\partial \xi_+}&\frac{1}{2}({\bf v}_{\pm}\nabla T)(\varepsilon_--\varepsilon_+)\left ( \xi_+ \frac{\partial n_+}{\partial \xi_+}+ \xi_- \frac{\partial n_-}{\partial \xi_-}\right )\\
\frac{1}{2}({\bf v}_{\mp}\nabla T )(\varepsilon_+-\varepsilon_-)\left ( \xi_+ \frac{\partial n_+}{\partial \xi_+}+ \xi_- \frac{\partial n_-}{\partial \xi_-}\right)&  ({\bf v}_{-}\nabla T)  \xi_- \frac{\partial n_-}{\partial \xi_-}            \end{array}\right)\nonumber\\
+               
 \left(
\begin{array} {cc}0&ig_{\pm}(\varepsilon_--\varepsilon_+)\\
ig_{\mp}(\varepsilon_+-\varepsilon_-)&0
 \end{array}\right)=I_{\alpha\beta}.~~~~~~~~~~~~~~~~~~~~~~~~~~~~~~~~~~~~~~~
 \label{eqv2}
\end{eqnarray}
\end{widetext}
The solution of Eq.(\ref{eqv2}) 
 has the following form
\begin{equation}
g_{\alpha\beta}= \left (\begin{array} {cc}g_+&g_{\pm}\\g_{\mp}&g_-
 \end{array}\right)=
 -\frac{1}{T}{\cdot}\left(
\begin{array} {cc} ({\bf u}_{+}\nabla T)
&({\bf u}_{\pm}\nabla T)
\\
({\bf u}_{\mp}\nabla T) 
&  ({\bf u}_{-} \nabla T)
            \end{array}\right).
 \label{sol2}
 \end{equation}
After substitution this matrix 
in the Eq.(\ref{eqv2}) and in the collision integral  (\ref{matrix1}) we obtain  four  equations corresponding to each matrix element
of  the matrix Eq.(\ref{sol2}) for four ${\bf k}$ dependent  scalar functions $ ({\bf u}_+{\nabla T})
,  
({\bf u}_\pm{\nabla T})
,
 ({\bf u}_\mp{\nabla T})
 , 
 ({\bf u}_-{\nabla T})
 $.
 
The  density of heat current  is
 \begin{equation}
 {\bf q}=\int \frac{d^3k}{(2\pi)^3}\xi_{\sigma\sigma_1}({\bf k})\frac{\partial \varepsilon_{\sigma_1\sigma_2}({\bf k})}{\partial {\bf  k}}g_{\sigma_2\sigma}({\bf k}).
\end{equation}
Transforming it to the band representation we obtain
\begin{eqnarray}
 {\bf q}=\int \frac{d^3k}{(2\pi)^3}
 \left \{\xi_+ {\bf v}_+g_+
 +\xi_-{\bf v}_-g_-\right.\nonumber\\
\left. +
 [\xi_+{\bf v}_\pm g_\mp-\xi_-{\bf v}_\mp g_\pm](\varepsilon_--\varepsilon_+)  \right \}
 .
\end{eqnarray}
 And substituting $g_{\alpha\beta}$ from Eq.(\ref{sol2})  we come to
\begin{eqnarray}
 {\bf q}=-\frac{1}{T}\int \frac{d^3k}{(2\pi)^3}\left \{ 
 \xi_+{\bf v}_+ ({\bf u}_{+}\nabla T)
 +\xi_-{\bf v}_-
 ({\bf u}_{-}\nabla T )\right.\nonumber\\
 \left.+ 
 \left [\xi_+{\bf v}_\pm({\bf u}_{\mp}\nabla T)
 -\xi_-{\bf v}_\mp({\bf u}_{\pm}\nabla T)
\right ](\varepsilon_--\varepsilon_+) 
 \right \}
 .
\end{eqnarray}

To find the heat current in the weak scattering regime one can use off-diagonal  matrix elements obtained in neglect of collisions
\begin{eqnarray}
g_\pm=-\frac{i}{2T}({\bf v}_{\pm}\nabla T)\left ( \xi_+ \frac{\partial n_+}{\partial \xi_+}+ \xi_- \frac{\partial n_-}{\partial \xi_-}\right ),\\
g_\mp=-\frac{i}{2T}({\bf v}_{\mp}\nabla T)\left ( \xi_+ \frac{\partial n_+}{\partial \xi_+}+ \xi_- \frac{\partial n_-}{\partial \xi_-}\right ).
\end{eqnarray}
Whereas the diagonal elements should be found from the equations
\begin{widetext}
\begin{eqnarray}
-\frac{1}{T}({\bf v}_{+}\nabla T)\xi_+ \frac{\partial n_+}{\partial \xi_+}=
4\pi n_i\int\frac{d^3k}{2\pi^3}|V({\bf k}-{\bf k}^\prime)|^2\times\nonumber~~~~~~~~~~~~~~~~~~~~~~~~~~~~~~~~~~~~~~~~~~~~~~
\\ \times\left\{
O_{++}({\bf k}{\bf k}^\prime)O_{++}({\bf k}^\prime{\bf k})[g_+({\bf k}^\prime)-g_+({\bf k})]\delta(\varepsilon_+^\prime-\varepsilon_+)+
O_{+-}({\bf k}{\bf k}^\prime)O_{-+}({\bf k}^\prime{\bf k})[g_-({\bf k}^\prime)-g_+({\bf k})]\delta(\varepsilon_-^\prime-\varepsilon_+)
\right \},~~\label{49}\\
-\frac{1}{T}({\bf v}_{-}\nabla T)\xi_- \frac{\partial n_-}{\partial \xi_-}=4\pi n_i\int\frac{d^3k}{2\pi^3}|V({\bf k}-{\bf k}^\prime)|^2\times\nonumber~~~~~~~~~~~~~~~~~~~~~~~~~~~~~~~~~~~~~~~~~~~~~~
\\
\times\left\{
O_{-+}({\bf k}{\bf k}^\prime)O_{+-}({\bf k}^\prime{\bf k})[g_+({\bf k}^\prime)-g_-({\bf k})]\delta(\varepsilon_+^\prime-\varepsilon_-)+
O_{--}({\bf k}{\bf k}^\prime)O_{--}({\bf k}^\prime{\bf k})[g_-({\bf k}^\prime)-g_-({\bf k})]\delta(\varepsilon_-^\prime-\varepsilon_-)
\right \}.~~\label{50}
\end{eqnarray}
\end{widetext}

\section{Electron-electron scattering}

The problem of electron-electron scattering in non-centrosymmetric metals has been discussed  in the paper \cite{Mineev2018}.
It was done making use the electron-electron collision integral  given by Eqs.(\ref{e27}), (\ref{e28}) for the spin-matrix distribution function derived by V.P.Silin \cite{Silin1971} and J.W.Jeon and W.J.Mullin \cite{Mullin1988} in application to the quasiparticles scattering in liquid $^3$He.  Giving the correct description of relaxation processes for spin-perturbed quasiparticle distributions in  Fermi liquid in a centrosymmetric media this integral  is not applicable for the description of relaxation in non-centrosymmetric case. Thus, the  approach developed in Ref.11 is not valid. 
The electron-electron  collision integral in a media without inversion center given by Eqs.(\ref{e27}),
(\ref{e29}) has much more cumbersome form. However, the main conclusion of Ref.11 is qualitatively correct: the zero temperature electron-electron scattering time in a non-centrosymmetric medium  is finite. 

In the equilibrium the integral (\ref{e27}),
(\ref{e29}) is equal to zero, but a deviation from equilibrium distribution results in non-vanishing collision terms even at zero temperature.
The reason for this  is that even at zero temperature the scattered quasiparticles can find the non-occupied states in between the two Fermi surfaces with different Fermi momenta corresponding two bands split by the spin-orbital coupling. The situation is in complete analog with spin-polarized liquid $^3$He, where
 the  scattering processes  for spin-diffusion  in transversal  to magnetic field direction involve all the states between  Fermi surfaces
 of  spin-up and spin-down quasiparticles, and the relaxation time acquires the finite zero-temperature value  \cite{Mullin1988,Meyerovich1990,Mineev2004}.
Appropriate also to mention the remark made by C.Herring \cite
{37} concerning a relaxation in ferromagnetics: "For a ferromagnetic metal\ldots.  if the spin of quasiparticle
at the Fermi surface is reversed, the corresponding quasiparticle
state will no longer be closed to the Fermi surface, and it will have a
finite, rather than an inifinitesimal, decay rate."  

The zero temperature decay rate causes a doubt in validity of the Fermi liquid approach to the description of electrons in
metals without inversion center. The estimation made in the Ref.11 and more careful calculations made for the polarised Fermi-gas \cite{10} allow to be sure in the applicability 
of the Fermi liquid theory so long
the splitting of Fermi surfaces in momentum space is small in comparison with the Fermi energy:
\begin{equation}
v_F(k_{F-}-k_{F+})\ll\varepsilon_F.
\end{equation}
The spin-orbital band splitting $v_F\Delta k_F$ is directly expressed through   the corresponding splitting of the de Haas - van Alphen magnetization oscillation frequencies \cite{Mineev2005}.
Determined experimentally the typical magnitude of band splitting  in many non-centrosymmetric metals is of the order of hundreds Kelvin \cite{Terashima2008, Onuki2014,Maurya2018}. This is much less than the Fermi energy.

Because of finite zero temperature electron-electron scattering relaxation time 
in a metal without inversion center the total resistivity at zero temperature consists of two 
 parts originating  from resistivity due to the electron-electron scattering  and 
due to the electron scattering on impurities
\begin{equation}
\rho=\rho_{ee}(T=0)+\rho_{imp}.
\end{equation}
We ignore here the tensorial character of resistivity.
The resistivity due impurity scattering is proportional to  impurity concentration $\rho_{imp}\propto n_{imp}$.
Thus, the zero-temperature resistivity due to electron-electron scattering $\rho_{ee}(T=0)$ can be experimentally found by the measuring of low temperature resistivity  at several finite impurity concentrations with subsequent taking the formal limit 
\begin{equation}
\rho_{ee}(T=0)=\rho (n_{imp} \to 0).
\end{equation}
The corresponding qualitative behaviour is shown in Fig.1.
Needless to say, the crystal should be practically  perfect because the presence of dislocations, tween boundaries and stacking faults  even in an almost ideally pure specimen can completely hide  a contribution of electron-electron scattering in the residual resistivity.
 The saturation of e-e contribution to resistivity
 has  been speculated to be related to the
 absence of usual $\propto T^2$ contribution in recent
experiments \cite{Peets2018}.

Similarly, the residual thermal resistivity consists of two parts determined by the scattering on impurities and the electron-electron scattering
\begin{equation}
\frac{T}{\kappa}=d_{imp}+d_{ee}(T=0).
\end{equation}

The ratio of thermal conductivity to resistivity at low temperatures in non-centrosymmetric materials  is still proportional to temperature according to the Wiedemann-Franz law
\begin{equation}
\frac{\kappa}{\sigma}=AT.
\end{equation}
However, the constant of proportionality  is  not universal Lorenz number but  acquires purity and substance dependent magnitude.

\section{Conclusion}

The spin-orbital coupling  in a medium without inversion centrum lifts the spin degeneracy of electronic states  and splits each conducting band in two bands with different Fermi momenta. The kinetic properties  in such  metal or  semiconductor
are   described by four kinetic equations for diagonal ( intra-band) and off-diagonal  (inter-band) matrix elements of distribution function. 
It is shown that the off-diagonal terms give rise the contribution in  charge, spin  and heat flows   not only due to the interband scattering 
but also  
 in the  collisionless case.
The theory of charge, spin and heat transport is drastically simplified at so called weak impurity scattering  when the impurity scattering rate does not exceed the energy of spin-orbit band splitting. This case one can use the collisionless solutions for the off-diagonal matrix elements of the distribution function and work with much simpler system of two kinetic equations for the diagonal elements.
The  derivations of  electron-impurity and electron-electron collision integrals  are presented.
Along with the scattering on impurities the electron-electron collisions in a medium without inversion center  are also responsible   for the finite zero temperature residual resistivity and residual thermal resistivity.

\appendix
\section{Electron-impurity collision integral}

The electron-impurity collision integral in the operator form \cite{Vasko2005} for spatially homogeneous  system is given by the formula
\begin{equation}
\hat I_{\sigma\sigma^\prime}=n_{imp} \int\frac{d^3q}{(2\pi)^3}|V({\bf q})|^2\int_{-\infty}^0d\tau e^{\lambda\tau}\hat A_{\sigma\sigma^\prime}(\tau,t)
\label{23}
\end{equation}
where $\lambda\to +0$ and the integrand in linear approximation in respect to slow varying in time density matrix 
$\rho_{\sigma_1\sigma_2}({\bf r},t)$ is 
\begin{eqnarray}
\hat A_{\sigma\sigma^\prime}(\tau,t)~~~~~~~~~~~~~~~~~~~~~~~~~~~~\nonumber\\=\left [ exp(i\hat h_{\sigma\sigma_1}\tau)[ e^{i{\bf q}{\bf r}}, \rho_{\sigma_1\sigma_2}({\bf r},t)]\exp(-i\hat h_{\sigma_2\sigma^\prime}\tau), e^{-i{\bf q}{\bf r}} \right ].~~~
\label{24}
\end{eqnarray}
 Square brakets means the commutator, 
$\sigma,\sigma_1,\dots$ are the spin indices, 
\begin{equation}
\hat h_{\sigma\sigma_1}=\hat\varepsilon_{\sigma\sigma_1}\left (-i\frac{\partial}{\partial {\bf r}}  \right )
\end{equation}
is the hamiltonian of noninteracting electrons  given by Eq.(\ref{H_0}) in coordinate representation. Its eigen functions satisfy  the equation
\begin{equation}
\hat h_{\sigma\sigma_1}e^{i{\bf k}{\bf r}}\Psi^\alpha_{\sigma_1}({\bf k})=\varepsilon_\alpha e^{i{\bf k}{\bf r}}\Psi^\alpha_\sigma({\bf k})
\label{26}
\end{equation}
In the Dirac notations they are $e^{i{\bf k}{\bf r}}\Psi^\alpha_{\sigma}({\bf k})=|{\bf k}\rangle\Psi^\alpha_{\sigma}({\bf k})$.
To transform collision integral from coordinate  to momentum representation and at the same time from spin to  
band representation one must  calculate the matrix element from the expression Eq.(\ref{24})
\begin{widetext}
\begin{eqnarray}
\Psi^{\alpha\star}_{\sigma}({\bf k})\langle{\bf k}|
A_{\sigma\sigma^\prime}(\tau,t)
|{\bf k}^\prime\rangle\Psi^\beta_{\sigma^\prime}({\bf k})=~~~~~~~~~~~~~~~~~~~~~~~~~~~~~~~~~~~\nonumber\\
=\int\frac{d^3k^\prime}{(2\pi)^3}\left \{\Psi^{\alpha\star}_{\sigma}({\bf k})\langle{\bf k}|
exp(i\hat h_{\sigma\sigma_1}\tau)
e^{i{\bf q}{\bf r}} \rho_{\sigma_1\sigma_2}({\bf r},t)
\exp(-i\hat h_{\sigma_2\sigma^\prime}\tau)|{\bf k}^\prime\rangle\Psi^\mu_{\sigma^\prime}({\bf k}^\prime)
~\Psi^{\mu\star}_{\sigma_3}({\bf k}^\prime)\langle{\bf k}^\prime| e^{-i{\bf q}{\bf r}}
|{\bf k}\rangle\Psi^\beta_{\sigma_3}({\bf k})\right. ~~~\nonumber\\
\left.-\Psi^{\alpha\star}_{\sigma}({\bf k})\langle{\bf k}|
exp(i\hat h_{\sigma\sigma_1}\tau)\rho_{\sigma_1\sigma_2}({\bf r},t)
e^{i{\bf q}{\bf r}} 
\exp(-i\hat h_{\sigma_2\sigma^\prime}\tau)|{\bf k}^\prime\rangle\Psi^\mu_{\sigma^\prime}({\bf k}^\prime)
~\Psi^{\mu\star}_{\sigma_3}({\bf k}^\prime)\langle{\bf k}^\prime| e^{-i{\bf q}{\bf r}}
|{\bf k}\rangle\Psi^\beta_{\sigma_3}({\bf k}) \right.
~~~\nonumber\\
\left.-\Psi^{\alpha\star}_{\sigma_3}({\bf k})\langle{\bf k}| e^{-i{\bf q}{\bf r}}
|{\bf k}^\prime\rangle
\Psi^\nu_{\sigma_3}
({\bf k}
^\prime )~
\Psi^{\nu\star}_{\sigma}({\bf k}^\prime)\langle{\bf k}|
exp(i\hat h_{\sigma\sigma_1}\tau)
e^{i{\bf q}{\bf r}} \rho_{\sigma_1\sigma_2}({\bf r},t)
\exp(-i\hat h_{\sigma_2\sigma^\prime}\tau)|{\bf k}\rangle\Psi^\beta_{\sigma^\prime}({\bf k})\right.
~~~\nonumber\\
\left.+\Psi^{\alpha\star}_{\sigma_3}({\bf k})\langle{\bf k}| e^{-i{\bf q}{\bf r}}
|{\bf k}^\prime\rangle
\Psi^\nu_{\sigma_3}
({\bf k}
^\prime )~
\Psi^{\nu\star}_{\sigma}({\bf k}^\prime)\langle{\bf k}|
exp(i\hat h_{\sigma\sigma_1}\tau)
 \rho_{\sigma_1\sigma_2}({\bf r},t)e^{i{\bf q}{\bf r}}
\exp(-i\hat h_{\sigma_2\sigma^\prime}\tau)|{\bf k}\rangle\Psi^\beta_{\sigma^\prime}({\bf k})\right \}
~~+ h.c.
\label{matel}
\end{eqnarray}
Let us use now the equation (\ref{26}) and the orthogonality conditions Eq.(\ref{ort}). For the first term in (\ref{matel}) we obtain
\begin{eqnarray}
\int\frac{d^3k^\prime}{(2\pi)^3}\Psi^{\alpha\star}_{\sigma}({\bf k})\langle{\bf k}|
exp(i\hat h_{\sigma\sigma_1}\tau)
e^{i{\bf q}{\bf r}} \rho_{\sigma_1\sigma_2}({\bf r},t)
\exp(-i\hat h_{\sigma_2\sigma^\prime}\tau)|{\bf k}^\prime\rangle\Psi^\mu_{\sigma^\prime}({\bf k}^\prime)
~\Psi^{\mu\star}_{\sigma_3}({\bf k}^\prime)\langle{\bf k}^\prime| e^{-i{\bf q}{\bf r}}
|{\bf k}\rangle\Psi^\beta_{\sigma_3}({\bf k})+h.c.~~~~~~\nonumber\\
=
\int\frac{d^3k^\prime}{(2\pi)^3}\Psi^{\alpha\star}_{\sigma}({\bf k})\langle{\bf k}|
\exp(i\varepsilon_\alpha\tau)
\Psi^{\nu}_{\sigma_1}({\bf k}^\prime)\Psi^{\nu\star}_{\sigma_3}({\bf k}^\prime)
e^{i{\bf q}{\bf r}} \rho_{\sigma_3\sigma_2}({\bf r},t)
\exp(-i\varepsilon_\mu^\prime\tau)|{\bf k}^\prime\rangle
\Psi^{\mu}_{\sigma_2}({\bf k}^\prime)\Psi^{\mu\star}_{\sigma_3}({\bf k}^\prime)
\delta({\bf k}^\prime+{\bf q}-{\bf k})
\Psi^\beta_{\sigma_3}({\bf k})+h.c.~~~~
\end{eqnarray}
Substituting this formula to Eq.(\ref{23}) 
 and performing the integration over $d^3q$ and over $\tau$ we come to
\begin{eqnarray} 
2\pi n_{imp}\int\frac{d^3k^\prime}{(2\pi)^3}|V({\bf k}-{\bf k}^\prime)|^2O_{\alpha\nu}({\bf k},{\bf k}^\prime)f_{ \nu\mu}({\bf k}^\prime)O_{\mu\beta}({\bf k}^\prime,{\bf k})
\delta(\varepsilon^\prime_\nu-\varepsilon_\beta), 
\end{eqnarray} 
where
\begin{equation}
f_{\alpha\beta}({\bf k})=\Psi^{\alpha\star}_{\sigma_1}({\bf k})\langle {\bf k}|\rho_{\sigma_1\sigma_2}({\bf r},t)|{\bf k}\rangle\Psi^{\beta}_{\sigma_2}({\bf k}),
\end{equation}
$ 
\varepsilon_\pm=\varepsilon_\pm({\bf k}),~
\varepsilon_\pm^\prime=\varepsilon_\pm({\bf k}^\prime)$ and
\begin{equation}
O_{\alpha\beta}({\bf k},{\bf k}^\prime)=\Psi^{\alpha\star}_\sigma({\bf k})\Psi^\beta_\sigma({\bf k}^\prime)
\end{equation}
such that
$
O_{\alpha\beta}({\bf k},{\bf k}^\prime)=O^\star_{\beta\alpha}({\bf k}^\prime,{\bf k}).
$
 Performing similar transformation with the other terms in Eq.(\ref{matel}) we come the  electron-impurity collision integral in band representation
 \begin{eqnarray}
I_{\alpha\beta}({\bf k})=2\pi n_{imp}\int\frac{d^3k^\prime}{(2\pi)^3}|V({\bf k}-{\bf k}^\prime)|^2\left \{O_{\alpha\nu}({\bf k},{\bf k}^\prime)\left [ f_{ \nu\mu}({\bf k}^\prime)O_{\mu\beta}({\bf k}^\prime,{\bf k})-O_{\nu\mu}({\bf k}^\prime,{\bf k})
  f_{ \mu\beta}({\bf k}) \right ]\delta(\varepsilon^\prime_\nu-\varepsilon_\beta)\right.\nonumber\\ 
 \left .+
  \left[O_{\alpha\nu}({\bf k},{\bf k}^\prime)f_{ \nu\mu}({\bf k}^\prime)-f_{ \alpha\nu}({\bf k})O_{\nu\mu}({\bf k},{\bf k}^\prime)
  \right ]O_{\mu\beta}({\bf k}^\prime,{\bf k})\delta(\varepsilon^\prime_\mu-\varepsilon_\alpha)\right \},
  \label{matrix}
\end{eqnarray}
\end{widetext}
The collision integral for a distribution function  varying  in space on the scales bigger than inter-electron distance has the same form.

One can rewrite the collision integral for the matrix distribution function as the integral depending from the vectorial distribution function
\begin{equation}
f^j({\bf k})=\frac{1}{2}f_{\alpha\beta}({\bf k})\sigma_{\beta\alpha}^j,
\end{equation}
where  $\sigma_{\beta\alpha}^j=(\delta_{\beta\alpha},\sigma^x_{\beta\alpha},\sigma^y_{\beta\alpha},\sigma^z_{\beta\alpha})$ is the four component matrix vector.
As result we obtain the scattering integral   amazingly similar to the standard diagonal in the spin indices expression
\begin{widetext}
\begin{eqnarray}
I_{\alpha\beta}({\bf k})=2\pi n_{imp}\int\frac{d^3k^\prime}{(2\pi)^3}|V({\bf k}-{\bf k}^\prime)|^2
\left \{O_{\alpha\nu}({\bf k},{\bf k}^\prime)R^j_{\nu\beta}({\bf k}^\prime,{\bf k})\delta(\varepsilon^\prime_\nu-\varepsilon_\beta)\right.
\nonumber\\
\left.+
R^j_{\alpha\nu}({\bf k},{\bf k}^\prime)O_{\nu\beta}({\bf k}^\prime,{\bf k})
\delta(\varepsilon^\prime_\nu-\varepsilon_\alpha)\right\}\left [f^j({\bf k}^\prime)-f^j({\bf k})\right ].
  \label{vector}
\end{eqnarray}
Here
\begin{eqnarray}
R^0_{\nu\beta}({\bf k}^\prime,{\bf k})=O^0({\bf k}^\prime,{\bf k})\delta_{\nu\beta}+
O^x({\bf k}^\prime,{\bf k})\sigma^x_{\nu\beta}  +
O^y({\bf k}^\prime,{\bf k})\sigma^y_{\nu\beta}  + O^z({\bf k}^\prime,{\bf k})\sigma^z_{\nu\beta},   \nonumber\\
R^x_{\nu\beta}({\bf k}^\prime,{\bf k})=O^x({\bf k}^\prime,{\bf k})\delta_{\nu\beta}+O^0({\bf k}^\prime,{\bf k})\sigma^x_{\nu\beta}  +
iO^y({\bf k}^\prime,{\bf k})\sigma^z_{\nu\beta}  -i O^z({\bf k}^\prime,{\bf k})\sigma^y_{\nu\beta},  \nonumber\\
R^y_{\nu\beta}({\bf k}^\prime,{\bf k}) =O^y({\bf k}^\prime,{\bf k})\delta_{\nu\beta}+
O^0({\bf k}^\prime,{\bf k})\sigma^y_{\nu\beta}  +
iO^z({\bf k}^\prime,{\bf k})\sigma^x_{\nu\beta}  -i O^x({\bf k}^\prime,{\bf k})\sigma^z_{\nu\beta} ,\nonumber\\
R^z_{\nu\beta}({\bf k}^\prime,{\bf k})=O^z({\bf k}^\prime,{\bf k})\delta_{\nu\beta}+
O^0({\bf k}^\prime,{\bf k})\sigma^z_{\nu\beta}  +
iO^x({\bf k}^\prime,{\bf k})\sigma^y_{\nu\beta}  -i O^y({\bf k}^\prime,{\bf k})\sigma^x_{\nu\beta}      ,
\end{eqnarray}
and $
O^j({\bf k}^\prime,{\bf k})=\frac{1}{2}O_{\alpha\beta}({\bf k})\sigma_{\beta\alpha}^j$.
\end{widetext}

\section{Electron-electron collision integral}

 The  Fermi particle-particle  collisions  integral in the Born approximation was derived  by V.P.Silin \cite{Silin1971} and by J.W.Jeon and W.J.Mullin \cite{Mullin1988} in application to liquid $^3$He.
In a crystal taking into account the Umklapp processes of scattering it is 
\begin{widetext}
\begin{equation}
\hat I({\bf k}_1)=2\pi\int d^3{\bf k}^\prime\frac{d^3{\bf k}^{\prime\prime}}{(2\pi)^3}\frac{d^3{\bf k}_2}{(2\pi)^3}\sum_{\bf m}
 \delta\left({\bf k}_1+{\bf k}_{2}-{\bf k}^\prime-{\bf
k}^{\prime\prime}-\frac{2\pi{\bf m}}{a}\right) \hat F,
\label{e27}
\end{equation}
where $\frac{2\pi{\bf m}}{a}$ is a vector of reciprocal lattice,
\begin{eqnarray}
\hat F
 =
 \left \{\frac{1}{2}W_1\left \{ [ \hat n^\prime,(\hat 1-\hat n_1) ]_+Tr((\hat 1-\hat n_2)n^{\prime\prime})-
 [(\hat 1- \hat n^\prime),\hat n_1 ]_+Tr (\hat n_2(\hat 1-\hat n^{\prime\prime}))\right \}\right.\nonumber\\
\left.+\frac{1}{2}
W_2\left\{[\hat n^\prime(\hat 1-\hat n_2)\hat n^{\prime\prime},(\hat 1-\hat n_1) ]_+-  [(\hat 1-\hat n^\prime)\hat n_2(\hat 1-\hat n^{\prime\prime}),\hat n_1 ]_+  \right \}\right \}\delta(\varepsilon_{1}+\varepsilon_{2}-\varepsilon^\prime-
\varepsilon^{\prime\prime}).
\label{e28}
\end{eqnarray}
\end{widetext}
Here $[\hat A,\hat B]_+$ means the anticommutator of the matrices $A$ and $B$, and the following designations $\hat n^\prime=\hat n({\bf k}^\prime),~ \varepsilon^\prime=\varepsilon ({\bf k}^\prime) $ etc are introduced.  In the isotropic Fermi liquid like $^3$He $W_1=[V(|{\bf k}_1-{\bf k}^\prime|)]^2,~~~~
W_2=-V(|{\bf k}_1-{\bf k}^\prime|)V(|{\bf k}_1-{\bf k}^{\prime\prime}|)$ are expressed trough the Fourier transform of the quasiparticles potential of interaction. The latter in concrete metal is unknown and due to  charge screening one can put them as the constants: $W_1=W_0/2,~~W_2=-W_0/2$.

This collision integral  can be also obtained
from the  integral of the electron-electron collisions in the operator representation
derived in Ref.23 where  was demonstrated that in the  case of diagonal in spin indices distribution function the collision integral  takes the standard form. 

The matrix $\hat F$ for the electron-electron collision integral corresponding to Eq. (\ref{e28})  for  the electron matrix distribution function in the band representation in non-centrosymmetric media is
\begin{widetext}
\begin{eqnarray}
F_{\alpha\beta}=\frac{1}{2}W_1\left \{  \left [O_{\alpha\nu}({\bf k}_1,{\bf k}^\prime) f_{ \nu\mu}({\bf k}^\prime)O_{\mu\lambda}({\bf k}^\prime,{\bf k}_1)(\delta_{\lambda\beta}-f_{\lambda\beta}({\bf k}_1))(\delta_{\xi\eta}-f_{\xi\eta}({\bf k}_2))O_{\eta\zeta}({\bf k}_2,{\bf k}^{\prime\prime}) f_{\zeta\rho}({\bf k}^{\prime\prime})O_{\rho\xi}({\bf k}^{\prime\prime},{\bf k}_2)\right.\right.~~~~~~~~~~~~~~~~~~~~\nonumber\\
-\left.\left.
O_{\alpha\nu}({\bf k}_1,{\bf k}^\prime)(\delta_{\nu\mu}- f_{ \nu\mu}({\bf k}^\prime))O_{\mu\lambda}({\bf k}^\prime,{\bf k}_1)
f_{\lambda\beta}({\bf k}_1)
f_{\xi\eta}({\bf k}_2)O_{\eta\zeta}({\bf k}_2,{\bf k}^{\prime\prime})(\delta_{\zeta\rho}- f_{\zeta\rho}({\bf k}^{\prime\prime}))O_{\rho\xi}({\bf k}^{\prime\prime},{\bf k}_2)\right ] 
\right.\delta(\varepsilon_\nu^\prime-\varepsilon_{1\beta} -\varepsilon_{2\xi} +\varepsilon_\zeta^{\prime\prime}) \nonumber\\
+
  \left.\left [(\delta_{\alpha\nu}-f_{\alpha\nu}({\bf k}_1))O_{\nu\mu}({\bf k}_1,{\bf k}^\prime)(\delta_{\mu\lambda}-f_{\mu\lambda}({\bf k}^\prime))O_{\lambda\beta}({\bf k}^\prime,{\bf k}_1)
  (\delta_{\xi\eta}-f_{\xi\eta}({\bf k}_2))O_{\eta\zeta}({\bf k}_2,{\bf k}^{\prime\prime}) f_{\zeta\rho}({\bf k}^{\prime\prime})O_{\rho\xi}({\bf k}^{\prime\prime},{\bf k}_2)\right.\right.~~~~~~~~~~~~~~~~~~~~ \nonumber\\
-
\left.\left.f_{\alpha\nu}({\bf k}_1))O_{\nu\mu}({\bf k}_1,{\bf k}^\prime)
f_{\mu\lambda}({\bf k}^\prime)O_{\lambda\beta}({\bf k}^\prime,{\bf k}_1)
f_{\xi\eta}({\bf k}_2)O_{\eta\zeta}({\bf k}_2,{\bf k}^{\prime\prime})(\delta_{\zeta\rho}- f_{\zeta\rho}({\bf k}^{\prime\prime}))O_{\rho\xi}({\bf k}^{\prime\prime},{\bf k}_2)\right ]\delta(\varepsilon_{1\alpha}-\varepsilon_\mu^{\prime}+\varepsilon_{2\xi} -\varepsilon_{\zeta}^{\prime\prime} )\right\} \nonumber\\
+ 
\frac{1}{2}W_2\left \{  \left [O_{\alpha\nu}({\bf k}_1,{\bf k}^\prime) f_{ \nu\mu}({\bf k}^\prime)O_{\mu\lambda}({\bf k}^\prime,{\bf k}_2)(\delta_{\lambda\varphi}-f_{\lambda\varphi}({\bf k}_2))O_{\varphi\psi}({\bf k}_2,{\bf k}^{\prime\prime})
f_{\psi\rho}({\bf k}^{\prime\prime}))
O_{\rho\omega}({\bf k}^{\prime\prime},{\bf k}_1)(\delta_{\omega\beta}- f_{\omega\beta}({\bf k}_1)~~~~~~~~~~~~~~~~~~~~\right.\right.\nonumber\\
-\left.\left.O_{\alpha\nu}({\bf k}_1,{\bf k}^\prime)(\delta_{\nu\mu}- f_{ \nu\mu}({\bf k}^\prime))O_{\mu\lambda}({\bf k}^\prime,{\bf k}_2)f_{\lambda\varphi}({\bf k}_2)O_{\varphi\psi}({\bf k}_2,{\bf k}^{\prime\prime})
(\delta_{\psi\rho}-f_{\psi\rho}({\bf k}^{\prime\prime}))
O_{\rho\omega}({\bf k}^{\prime\prime},{\bf k}_1)f_{\omega\beta}({\bf k}_1\right ] \delta(\varepsilon_\nu^\prime-\varepsilon_{1\beta} -\varepsilon_{2\varphi} +\varepsilon_\psi^{\prime\prime})\right.\nonumber\\
+ 
\left.  \left [(\delta_{\alpha\nu}- f_{\alpha\nu}({\bf k}_1)O_{\nu\mu}({\bf k}_1,{\bf k}^\prime) f_{ \mu\lambda}({\bf k}^\prime)O_{\lambda\varphi}({\bf k}^\prime,{\bf k}_2)(\delta_{\varphi\psi}-f_{\varphi\psi}({\bf k}_2))O_{\psi\rho}({\bf k}_2,{\bf k}^{\prime\prime})
f_{\rho\omega}({\bf k}^{\prime\prime}))
O_{\omega\beta}({\bf k}^{\prime\prime},{\bf k}_1)~~~~~~~~~~~~~~~~~~~~~~~~~\right.\right.\nonumber\\
-
\left.  \left. f_{\alpha\nu}({\bf k}_1)O_{\nu\mu}({\bf k}_1,{\bf k}^\prime)(\delta_{\mu\lambda}- f_{ \mu\lambda}({\bf k}^\prime))
O_{\lambda\varphi}({\bf k}^\prime,{\bf k}_2)f_{\varphi\psi}({\bf k}_2)O_{\psi\rho}({\bf k}_2,{\bf k}^{\prime\prime})
(\delta_{\rho\omega}-f_{\rho\omega}({\bf k}^{\prime\prime}))
O_{\omega\beta}({\bf k}^{\prime\prime},{\bf k}_1)\right]\delta(\varepsilon_{1\alpha}-\varepsilon_{\mu}^\prime +\varepsilon_{2\psi} -\varepsilon_\rho^{\prime\prime})\right\}.
\nonumber\\
 \label{e29}
\end{eqnarray}

\end{widetext}

\begin{figure}[p]
\includegraphics
[height=.8\textheight]
{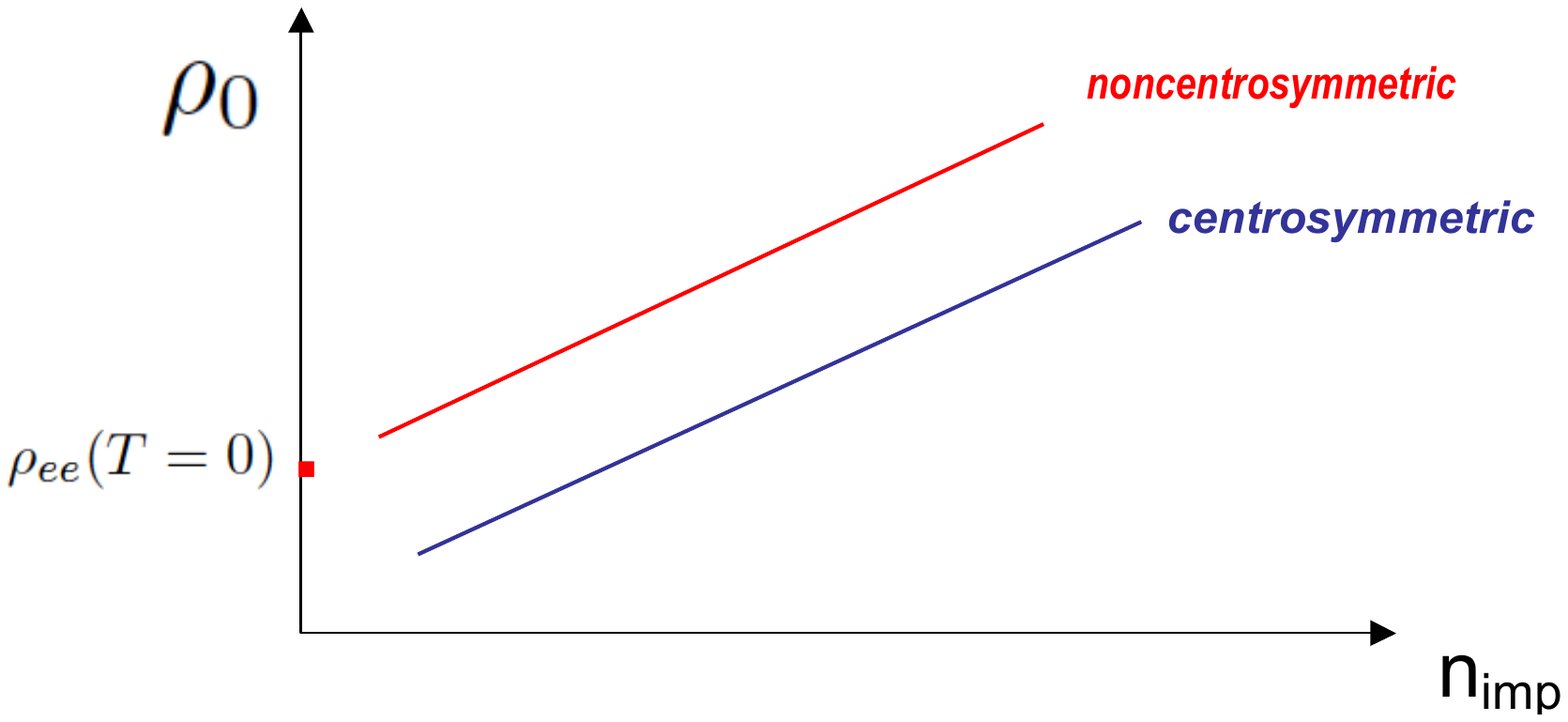}
 \caption{(Color online) 
 Residual resistivity }
\end{figure}

\end{document}